\documentclass[aps,prl,superscriptaddress,nofootinbib,twocolumn,10pt]{revtex4-1}
\usepackage[utf8]{inputenc}
\usepackage{natbib}
\usepackage{amsmath}
\usepackage{amsbsy}
\usepackage{amssymb}
\usepackage{scrextend}
\usepackage{graphicx}
\usepackage{color}
\usepackage{xcolor}

\usepackage{hyperref}

\graphicspath{{graphics/}}

\newcommand{\SBS}{\mathrm{Sr}_{0.1}\mathrm{Bi}_{2}\mathrm{Se}_{3}}
\renewcommand{\vec}[1]{\boldsymbol{#1}}

\newcommand{\supercite}[1]{$^{\text{\cite{#1}}}$}
\begin{document}
\title{Nanocalorimetric Evidence for Nematic Superconductivity in the\\Doped Topological Insulator Sr$_{0.1}$Bi$_{2}$Se$_{3}$}
%
%
\author{Kristin Willa}
\email[Corresponding author: ]{kwilla@anl.gov}

\affiliation{Materials Science Division, Argonne National Laboratory, Lemont IL 60439, USA}
\author{Roland Willa}
\affiliation{Materials Science Division, Argonne National Laboratory, Lemont IL 60439, USA}
%
\author{Kok Wee Song}
\affiliation{Materials Science Division, Argonne National Laboratory, Lemont IL 60439, USA}
\author{G. D. Gu}
\affiliation{Condensed Matter Physics and Materials Science Department, Brookhaven National Laboratory, Upton NY 11793, USA}
\author{John A.\ Schneeloch}
\affiliation{Condensed Matter Physics and Materials Science Department, Brookhaven National Laboratory, Upton NY 11793, USA}
\affiliation{Department of Physics and Astronomy, Stony Brook University, Stony Brook NY 11794, USA}
\author{Ruidan Zhong}
\affiliation{Condensed Matter Physics and Materials Science Department, Brookhaven National Laboratory, Upton NY 11793, USA}
\affiliation{Department of Materials Science and Engineering, Stony Brook University, Stony Brook NY 11794, USA}
\author{Alexei E.\ Koshelev}
\affiliation{Materials Science Division, Argonne National Laboratory, Lemont IL 60439, USA}
\author{Wai-Kwong Kwok}
\affiliation{Materials Science Division, Argonne National Laboratory, Lemont IL 60439, USA}
\author{Ulrich Welp}
\affiliation{Materials Science Division, Argonne National Laboratory, Lemont IL 60439, USA}
\date{\today}

\begin{abstract}
Spontaneous rotational-symmetry breaking in the superconducting state of doped $\mathrm{Bi}_2\mathrm{Se}_3$ has attracted significant attention as an indicator for topological superconductivity. In this paper, high-resolution calorimetry of the single-crystal $\mathrm{Sr}_{0.1}\mathrm{Bi}_2\mathrm{Se}_3$ provides unequivocal evidence of a two-fold rotational symmetry in the superconducting gap by a \emph{bulk thermodynamic} probe, a fingerprint of nematic superconductivity.  The extremely small specific heat anomaly resolved with our high-sensitivity technique is consistent with the material's low carrier concentration proving bulk superconductivity. The large basal-plane anisotropy of $H_{c2}$ is attributed to a nematic phase of a two-component topological gap structure $\vec{\eta} = (\eta_{1}, \eta_{2})$ and caused by a symmetry-breaking energy term $\delta (|\eta_{1}|^{2} - |\eta_{2}|^{2}) T_{c}$. A quantitative analysis of our data excludes more conventional sources of this two-fold anisotropy and provides the first estimate for the symmetry-breaking strength $\delta \approx 0.1$, a value that points to an onset transition of the second order parameter component below 2K.
\end{abstract}
\maketitle
The prospect of fault-tolerant quantum computing based on the non-Abelian braiding properties of Majorana fermions has generated enormous interest in the synthesis and study of topological superconductors\supercite{Nayak2008, Merali2011, Qi2009, Qi2011}. Currently, two paths towards topological superconductivity are being pursued: proximity-induced topological states at the \emph{interface} between a conventional superconductor and a topological insulator or a strong spin-orbit coupled semiconductor\supercite{Fu2008, Mourik2012, Albrecht2016}, respectively, or by doping-induced superconductivity in \emph{bulk} topological insulators\supercite{Sasaki2015, Sato2017}. Among the doped topological insulators, $M\mathrm{Bi}_{2}\mathrm{Se}_{3}$ (with $M \!=\! \mathrm{Cu}$\supercite{Hor2010}, $\mathrm{Nb}$\supercite{Qiu2015-condmat}, $\mathrm{Sr}$\supercite{Liu2015}) have attracted considerable interest since they display phenomenology---the spontaneous emergence of a two-fold in-plane anisotropy of various superconducting quantities in a three-fold in-plane crystal structure\supercite{Matano2016, Yonezawa2016, Asaba2017, Shen2017} and evidence for a nodal gap\supercite{Smylie2016, Smylie2017a}---that is consistent with a topological state. 
Several theoretical works\supercite{Nagai2012, Fu2014, Venderbos2016a} propose a two-component superconducting order parameter $\vec{\eta} = (\eta_{1}, \eta_{2})$   of $E_u$ symmetry for this class of materials. This order parameter by itself does not lead to a two-fold in-plane symmetry; yet the asymmetric coupling of the two order parameter components, analogous to the case of the unconventional superconductor $\mathrm{UPt}_{3}$\cite{Joynt2002}, causes a symmetry breaking. For the doped $\mathrm{Bi}_2\mathrm{Se}_3$, a coupling to the strain field, $\delta (|\eta_{1}|^{2} - |\eta_{2}|^{2}) T_{c}$, has been proposed to induce the nematic state, where the coupling strength is quantified by the phenomenological parameter $\delta$ \supercite{Venderbos2016a}.
\begin{figure}[tb]
\centering
\includegraphics[width=0.48\textwidth]{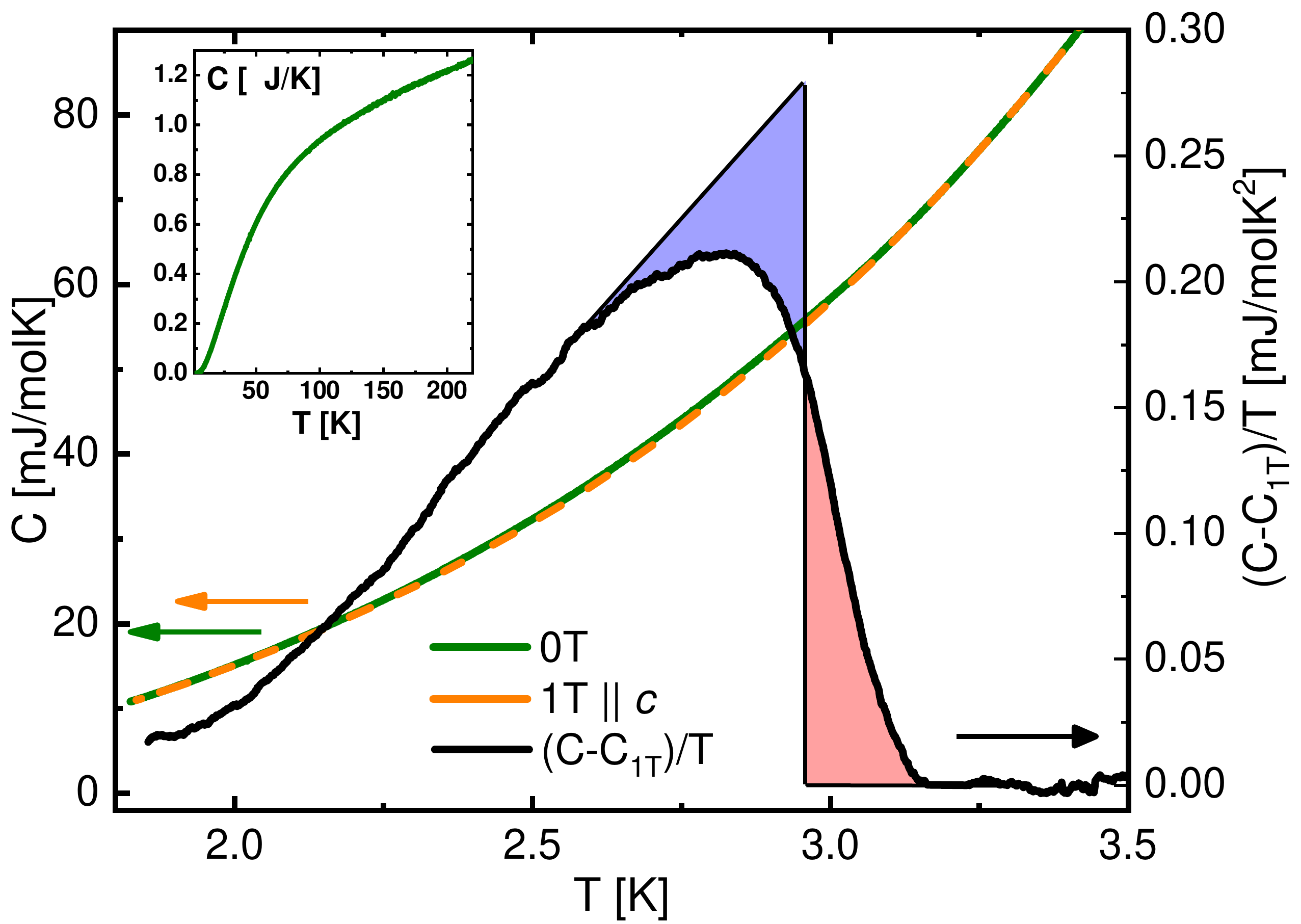}
\caption{Zero-field heat capacity measurement of $\SBS$ from room temperature down to $1.8\mathrm{K}$ (green, inset) and a close-up near the superconducting transition (main figure). The latter becomes visible only after subtraction of a normal-state background (orange). The subtracted curve (black, and normalized by $T$) reveals a small relative specific heat jump, $\Delta C / C \sim 10^{-2}$; yet compatible with bulk superconductivity.
}
\label{fig:sc-transition}
\end{figure}

The two-fold $H_{c2}$-anisotropy in Sr$_{x}$Bi$_{2}$Se$_{3}$ has been studied using several complementary approaches, including magnetotransport\supercite{Pan2016}, transport under pressure\supercite{Nikitin2016}, Corbino geometry\supercite{Du2017}, and magnetization\supercite{Smylie2018b}. While confirming the two-fold rotational anisotropy, these studies also point towards an isotropic response in the \textit{normal state}, i.e., excluding conventional sources of anisotropy such as an elliptic Fermi surface, structural inhomogeneities, or magnetic impurities. A specific heat study\supercite{Pan2016} concluded that no structural transition is breaking the crystalline symmetry. However, this measurement did not yield a discernible specific heat anomaly at the superconducting transition. Whereas the two-fold symmetry has been demonstrated beyond any doubt, pressing open questions remain, e.g., on the selection of the nematic direction and the origin of the strain field. Determining the strength of this field $\delta$ is essential in understanding the nematic state in these compounds and thus an important step on the way to find topological superconductivity. 

In this paper, we report calorimetric measurements on Sr$_{0.1}$Bi$_{2}$Se$_{3}$ single crystals synthesized by the melt-growth technique\cite{Du2017} and show the first observation of a clear step of 0.28 mJ/mol\ K$^2$ in the specific heat at the superconducting transition near ~3K.  The step height, albeit small, is in agreement with estimates based on magnetization measurements and the electronic band structure.  The specific heat measurements do not reveal a double-transition as seen for instance in UPt$_3$\cite{Joynt2002} establishing boundaries on the coupling strength to a symmetry-breaking strain field of $\delta  >  0.1$. In conjunction with a theoretical analysis of the upper critical field of a superconductor with a two-component order parameter our measurements reveal that Sr$_{0.1}$Bi$_{2}$Se$_{3}$ is in the strong-coupling regime accompanied by a temperature-independent in-plane anisotropy.

The $ac$ specific heat is measured on a $\mathrm{Si}\mathrm{N}$ calorimetric membrane\supercite{Tagliati2012, Willa2017}, and the experiment is controlled with a SynkTek MCL1-540 multi-channel lock-in system. A small ($200 \times 300 \times 25 \mu\mathrm{m}^{3}$) platelet-shaped single crystal is mounted on the nanocalorimeter platform with apiezon grease, and inserted into a 1-1-9T three axis superconducting vector magnet. The $c$ axis of the sample is aligned with the magnet's 9T (or $z$-) direction.

\begin{figure*}[tbh!]
\includegraphics[width=0.93\linewidth]{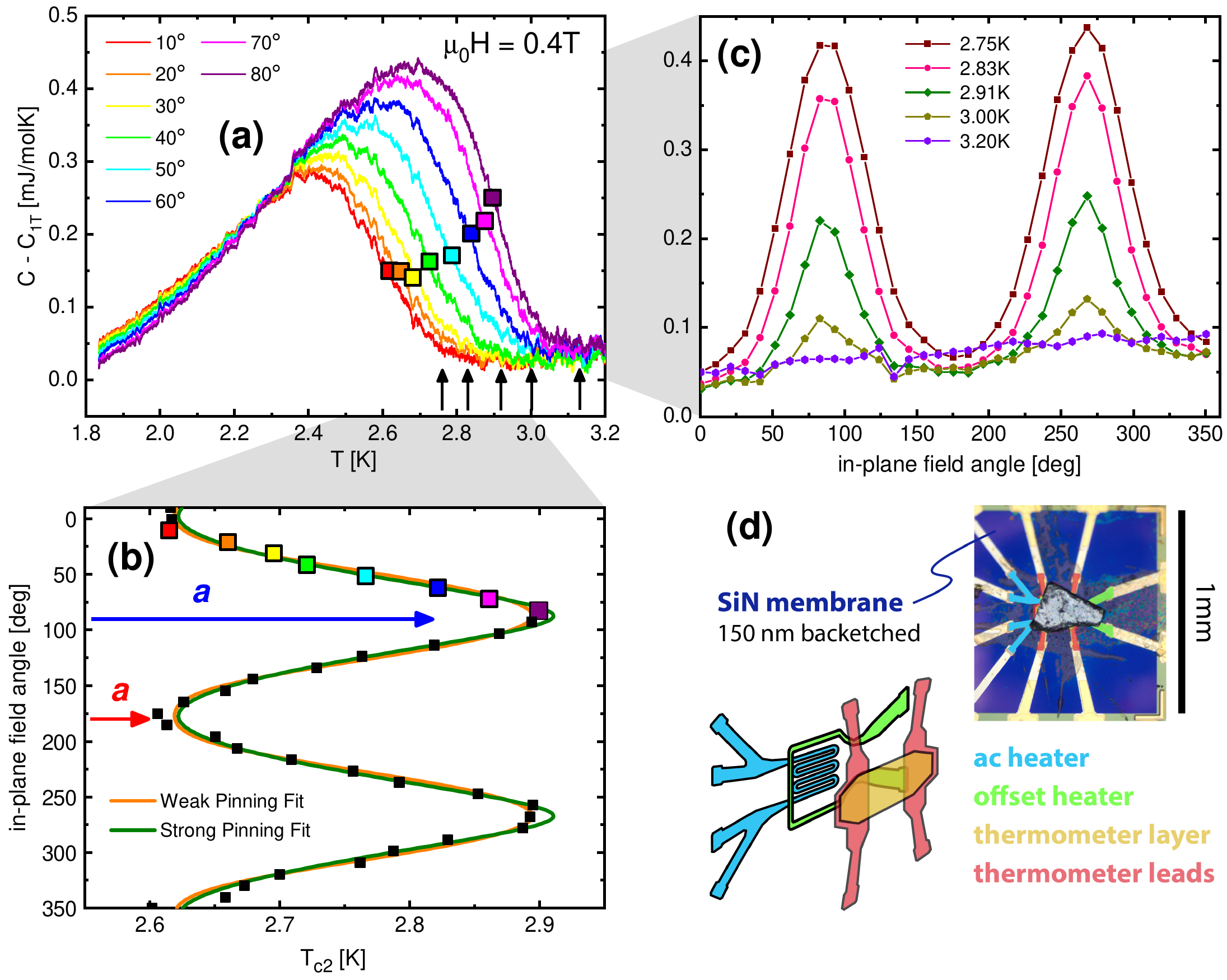}
\caption{
\textbf{(a)} Calorimetric scans---taken at fixed magnetic field strength $H = 0.4\mathrm{T}$---for different field orientations in the basal plane.
\textbf{(b)} Angular dependence of the upper critical temperature $T_{c2}(\theta)$, where each point results from a full specific heat scan [see colored squares in (a)]. $T_{c2}(\theta)$ assumes a maximal (minimal) value $2.9\mathrm{K}$ ($2.6\mathrm{K}$) along the $a$ ($a^{\star}$) direction. Weak and strong pinning fits are discussed in the main text below.
\textbf{(c)} Appearance of anisotropy in the specific heat when the sample is cooled through the superconducting transition. Each curve corresponds to a cut at constant temperature, [indicated by vertical arrows in (a)]. The absence of an anisotropy above $T_{c}$ is consistent with earlier works on magnetization and magnetotransport.
\textbf{(d)} Microscope image of $\SBS$ single crystal on the nanocalorimetric platform, and platform architecture.
}
\label{fig:Tc-of-theta}
\end{figure*}

The measured heat capacity (chip background subtracted) is featureless and does not reveal any indication for a structural transition occurring during cool-down from room temperature (Figure \ref{fig:sc-transition}). Even the superconducting transition is not apparent in the raw data (see main panel). However, the superconducting transition becomes visible after subtracting normal-state data $C_{1\mathrm{T}}/T$ (obtained by applying $1\mathrm{T}$ along the $c$ axis, see Reference [\onlinecite{Smylie2018b}]) from the zero-field curve. The transition temperature---as extracted from an entropy conserving construction---amounts to $T_{c} = 2.95\mathrm{K}$. The step height at the transition of $0.28\ \mathrm{mJ}/\mathrm{mol}\,\mathrm{K}^{2}$ corresponds to about $1\%$ of the total signal and is $\sim 50$ times smaller than in a conventional superconductor (e.g.\ lead).
Superconducting shielding fractions of 70\% have been reported \cite{Leng2018} while thermodynamic evidence was missing so far. We now want to verify the consistency of the measured magnitude of $\Delta C$ with other superconducting and normal state parameters of the material to show bulk superconductivity.
The relation $\Delta C = (T_{c} / 4 \pi) (\partial H_{c} / \partial T )^2 |_{T_c}$---originally derived by Rutgers\supercite{Rutgers1934}---provides an estimate for the jump in the specific heat in terms of the slope of the critical field $H_{c}$. Here, $\Delta C$ is given in $\mathrm{erg}\, \mathrm{cm}^{-3} \mathrm{K}^{-1}$. Substituting $H_{c}=H_{c2}/\kappa\sqrt 2$ and using the relation $\partial M/\partial T = - (8 \pi \beta_{A} \kappa^{2})^{-1} ( \partial H_{c2} / \partial T)_{T_{c}}$ [with $\beta_{A} \approx 1.16$] derived\supercite{deGennes1966} for a large Ginzburg-Landau parameter $\kappa$, we arrive at
\begin{align}\label{}
   \Delta C / T_{c} = - \beta_{A} \big(\partial M / \partial T \big) \big( \partial H_{c2} / \partial T\big)_{T_{c}} 
\end{align}
These relations remain valid for a multi-component order parameter in the linear regime near T$_c$. With the slope $\partial M / \partial T = 2.5\ 10^{-3} \mathrm{emu}/ \mathrm{cm}^{3} \mathrm{K}$ reported in Reference [\onlinecite{Smylie2018b}] for single crystals from the same synthesis batch] and the slope $\partial H_{c2}/\partial T = -10 \mathrm{kG}/\mathrm{K}$ determined below, we arrive at an estimate $\Delta C/ T_{c} = 0.2\mathrm{mJ}/\mathrm{mol}\,\mathrm{K}^{2}$ which agrees well with the observed value and demontrates that different measured parameters are thermodynamically consistent.
Within single-band weak-coupling BCS theory\supercite{Bardeen1957}, the jump in the specific heat satisfies the relation $\Delta C/ T_{c} =  1.43 \gamma$, with the Sommerfeld coefficient $\gamma = (\pi^{2}/2) k_{B}^{2} n / \epsilon_{F}$ expressed in terms of the charge carrier density $n$ and the Fermi energy $\epsilon_{F}$. We use reported values for $n$ and $\epsilon_{F}$ obtained from Seebeck and Hall effect measurements\supercite{Shruti2015} to evaluate the Sommerfeld coefficient. These estimates provide a value for $\Delta C/T_{c} = 0.42\ \mathrm{mJ/molK}^{2}$, again close to our measured value. With these estimates, we exclude the scenario of a tiny superconducting volume fraction.

\begin{figure*}[tbh]
\centering
\includegraphics[width=.98\linewidth]{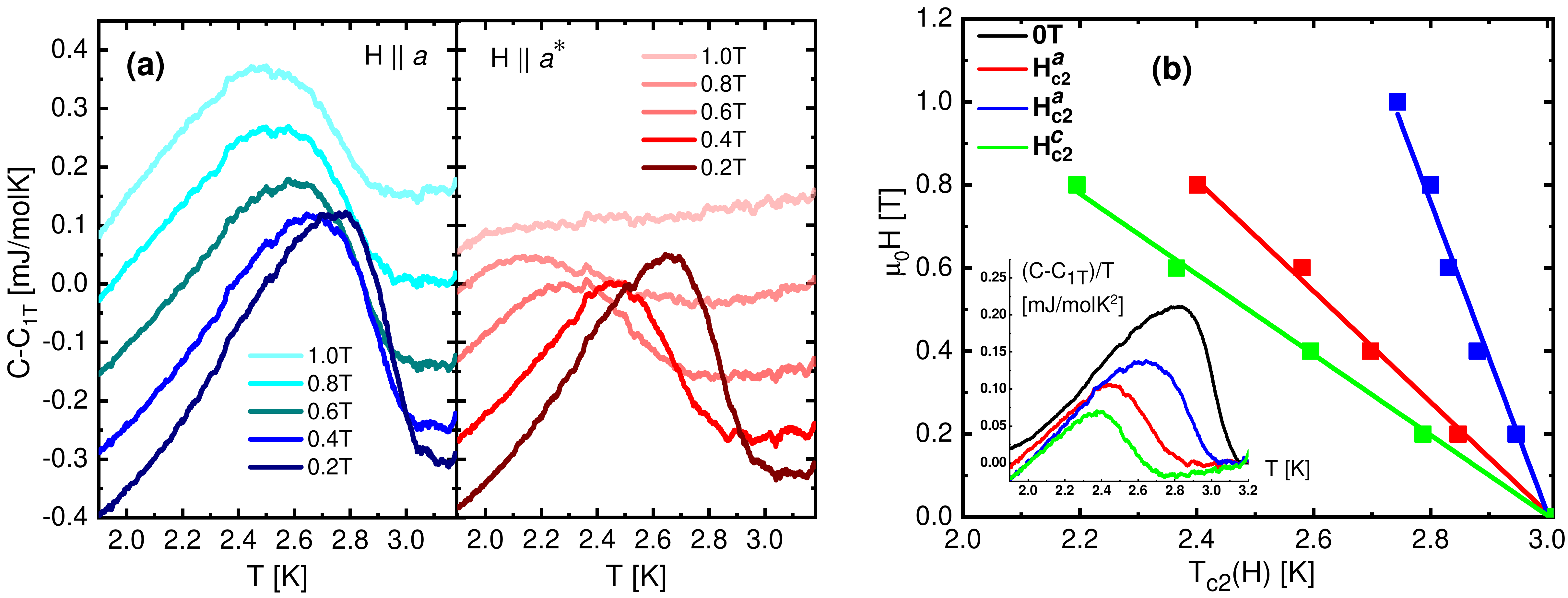}
\caption{\textbf{(a)} Calorimetric scans along the basal principal axes for various field strengths (curves are arbitrarily offset for better visibility), see blue and red arrows in Figure \ref{fig:Tc-of-theta} (b), reveal a robust superconducting order along $a$ and a less robust one along $a^{\star}$. \textbf{(b)} The $H_{c2}$ phase boundary---shown for all three principal axes---features a temperature-independent anisotropy $\Gamma_{\mathrm{exp}} = 3.5$ between the two basal-plane upper critical fields. A comparison between the 0.4T calorimetric scans along the three crystallographic directions (blue, red, green) and the zero field specific heat (black) is given in the inset.
}
\label{fig:Hmin-Hmax}
\end{figure*}
The in-plane anisotropy of the specific heat is studied by applying a field of fixed strength ($0.4$ Tesla) in the plane normal to the crystallographic $c$ axis and by rotating it in steps of 10 degrees. At each step we measure the specific heat upon cooling through the transition from 3.2K down to 1.8K. A set of scans is shown in Figure \ref{fig:Tc-of-theta} [with the normal state (1T along the $c$ axis) curve already subtracted]. The transition temperature and magnitude clearly depend on the in-plane field orientation. For each field angle, the upper critical temperature $T_{c2}(\theta)$ is extracted from the inflection point%
\footnote{It shall be noted here, that a more rigorous definition of the transition temperature involves an entropy conserving construction. The latter however requires an accurate determination of the specific heat away from the transition temperature, a criterion that is not systematically met for all angles. Comparing both methods (where applicable) reveals a small shift by $\sim 0.05\mathrm{K}$ in $T_{c}$ and a negligible difference in its angular dependence.}
in $[C(T) - C_{1\mathrm{T}}(T)]/T$, see Figure \ref{fig:Tc-of-theta}(a). As a function of $\theta$, the upper critical temperature shows a strong twofold in-plane anisotropy, with $T_{c2}^{\mathrm{max}} = 2.9\mathrm{K}$ and $T_{c2}^{\mathrm{min}} = 2.6\mathrm{K}$ along two directions separated by $90^{\circ}$. These axes have previously been identified as the crystallographic $a$ and $a^{*}$ directions, respectively. \cite{Pan2016, Smylie2018b}
Focusing on the principal axes, we performed more detailed measurements of the superconducting phase boundary with field strengths between 0.2T and 1T. From the specific heat curves along the two extremal in-plane directions, shown in panel (a) of Figure \ref{fig:Hmin-Hmax},
the $H_{c2}$ phase boundary can be determined, see Figure \ref{fig:Hmin-Hmax}(b). The in-plane anisotropy of $H_{c2}$ amounts to $\Gamma_{\mathrm{exp}} = H_{c2}^{a}/H_{c2}^{a^{*}} = 3.5$ and is \emph{independent} of the temperature (within the studied field range). For completeness the $c$-axis phase boundary is also determined; a direct comparison of the calorimetric curves for these three directions (for $0.4$ Tesla) is shown in Figure \ref{fig:Hmin-Hmax}(b).
We note that in the normal state the specific heat is isotropic in the plane as evidenced by the 3.2K-data in Figure \ref{fig:Tc-of-theta}(c).
We thus exclude normal-state properties as the cause of in-plane anisotropy of $H_{c2}$. An alternative explanation for an anisotropic $H_{c2}$ in doped $\mathrm{Bi}_{2}\mathrm{Se}_{3}$ was laid out by Venderbos and co-workers\supercite{Venderbos2016a} (earlier work on $\mathrm{U}\mathrm{Pt}_{3}$ goes back to Agterberg \emph{et al.}\supercite{Agterberg1995, Agterberg1995b}) and invokes a two-component order parameter $\vec{\eta} = (\eta_{1},\eta_{2})$. The order parameter is usually treated by the Ginzburg-Landau (GL) formalism, where the linearized GL equations near $H_{c2}$ read
\begin{align}\label{eq:Venderbos}
   \!(T_{c0}-T) \eta_{a} &= J (D_{x}^{2} + D_{y}^{2})\eta_{a} + K D_{z}^{2}\eta_{a} - \delta T_{c0} \tau_{3}^{ab} \eta_{b}\!\\\nonumber
   &\mu J [(D_{x}^{2} - D_{y}^{2})\tau_{3}^{ab} + \ (D_{x}D_{y}-D_{y}D_{x})\tau_{1}^{ab}]\eta_{b},
\end{align}
with $T_{c0}$ the bare transition temperature (when $\delta = 0$), $D_{\alpha} = -i\partial_{\alpha} - 2eA_{\alpha}$ ($\alpha = x,y,z$) are gauge-invariant gradients, with $\vec{A}$ the electromagnetic vector potential, and $\tau_{\alpha}$ are the Pauli matrices, \{\} the anticommutator, and $\hbar=1$. Summation is over double indices. In this picture, a two-fold anisotropy can exists only if both parameters $\delta$ (coupling the order parameters to the strain), and $\mu$ (ratio between GL parameters for the isotropic and mixed gradient terms) are non-vanishing. A finite $\delta$ (we take $\delta > 0$ for definition) shifts the mean-field transition temperature of the component $\eta_{1}$ to $T_{c} \equiv T_{c0}(1 + \delta)$ and that of $\eta_{2}$ to $T_{c0}(1 - \delta)$ respectively.
It takes a substantial effort and limiting assumptions to derive the angular dependence of $H_{c2}(\theta)$. Several such limits have been considered in preceding works\supercite{Venderbos2016a, Shen2017}. However, the anisotropy ratio $\Gamma = H_{c2}^{\mathrm{max}}/H_{c2}^{\mathrm{min}}$ between the maximal and minimal field directions can be solved without any restriction on the magnitude of $\delta$ and $\mu$, see Supplemental Material \cite{suppl}. It turns out that for sufficiently large $\delta > \delta_{c}$, with $\delta_{c}$ determined by $2\delta_{c}/(1+\delta_{c}) = (1 - T/T_{c})\{1-[(1-\mu)/(1+\mu)]^{1/2}\}$, the anisotropy ratio reads $\Gamma_{>} = [(1+\mu)/(1-\mu)]^{1/2}$ independent of both $\delta$ and the temperature. Below that bound, the ratio reads $\Gamma_{<} = 1 + 2\delta / [(1+\delta)(1-T/T_{c})-2\delta]$, independent of $\mu$ \emph{and} explicitly temperature-dependent.

Our experiment clearly indicates a temperature-independent anisotropy, hence excluding the scenario for very small $\delta \ll \delta_{c}$. An implicit equation for $H_{c2}(\theta)$ has been derived for moderate $\delta$ ($\lesssim 1-\mu$), see Equation (108) in the Supplementary Materials of Reference [\onlinecite{Venderbos2016a}]. A numerical fit (weak pinning fit, see Figure \ref{fig:Tc-of-theta}) to our data shows a very good agreement, see Figure \ref{fig:Tc-of-theta}(b), and provides an estimate for $\mu = 0.82$ and $\delta = 0.09$. These values imply that the appearance of the second (suppressed) order parameter near $T_{c,2} \approx T_{c} (1-\delta)/(1+\delta) \sim 2.5 \mathrm{K}$ causes a second discontinuity in the specific heat; a feature that is not resolved in the data.  

For UPt$_3$---an extensively studied unconventional heavy-Fermion superconductor with a two-component order parameter \cite{Joynt2002}---it is believed that weak antiferromagnetism lifts the degeneracy of the order parameter components giving rise to two distinct zero-field transitions in the specific heat split apart by ~60 mK \cite{Fisher1989, Bogenberger1993, Jin1992, Hasselbach1989}. A phenomenological GL analysis \cite{Sigrist1991} shows that the ratio of the specific heat jumps at the two transition temperatures ($T_{c}$ and $T_{c,2}$) involves the phenomenological parameters associated to the quartic terms in the GL free energy. As a result, no parametric smallness is imposed on the second calorimetric discontinuity. A thermodynamic analysis \cite{Yip1991, Boukhny1994} reveals that the amplitude ratio of the specific heat anomalies depends on the slopes of the phase boundaries separating the normal state and the various order parameter configurations, respectively. Here again, unless these boundaries are very steep or horizontal, the amplitudes of the specific heat anomalies are expected to be of the same order of magnitude.
In the case of Sr$_{0.1}$Bi$_{2}$Se$_{3}$, only the normal state boundary is currently known. Noting the absence of a second discontinuity in the specific heat, our results therefore imply that $\delta$ is large, shifting the second transition to low temperatures. Then, by identifying the measured anisotropy of 3.5 with $\Gamma_{>}$, a value of $\mu$ = 0.85 and a lower bound for $\delta > \delta_{c} \approx 0.11$ are obtained. For very large $\delta \gg \delta_{c}$, the order parameter is pinned to the pure form $\vec{\eta} = (\eta_{1}, 0)$ at $H_{c2}$ (along any angle), and a simple effective-mass like dependence $H_{c2}^{\mathrm{max}} [\cos^{2}\vartheta + \Gamma^{2} \sin^{2}\vartheta]^{-1/2}$ can be derived\supercite{Shen2017} (here $\vartheta \equiv \theta - \theta_{\mathrm{max}}$ is the angle measured away from the maximal $H_{c2}$-direction). A corresponding (strong pinning) fit to our experimental result is shown in Figure \ref{fig:Tc-of-theta}(b) producing the same fit quality as for the weaker pinning field.

We have investigated the anisotropic response of the superconducting state of $\SBS$ through high-precision calorimetric measurements. From our work we conclude that (i) the normal state has an isotropic basal plane, (ii) no structural transition is observed down to 1.8K, (iii) the jump in the specific heat at $T_{c}$ is small, consistent with the behavior of the magnetization and with a very low electron concentration, and (iv) the basal-plane anisotropy of $H_{c2}$ is large, i.e. $\Gamma_{\mathrm{exp}} = 3.5$ and temperature-independent. The prevailing theoretical explanation for this anisotropy is based \supercite{Venderbos2016a} on a two-component gap function realizing a nematic superconducting state due to possibly strain-induced symmetry-breaking. Within this framework, the experimental data allows (v) to estimate the ratio $\mu = 0.85$ between the GL parameters for the isotropic and mixed gradient terms, and (vi) most importantly provides a lower bound for the symmetry-breaking strength $\delta > \delta_{c}\approx 0.1$, the parameter that causes the decoupling of the two order parameters. 

Whereas the appearance of the order parameter's second component is expected to leave its trace in calorimetry, see References [\onlinecite{Fisher1989}] and [\onlinecite{Joynt2002}] on $\mathrm{U}\mathrm{Pt}_{3}$, we surmise this feature to be below 1.8K giving another indication of a strong pinning field $\delta$. These results open the door for further studies looking for the proof to topological superconductivity in the material class of the doped Bi$_2$Se$_3$.

\begin{acknowledgments}
We cordially thank Matthew P.\ Smylie for fruitful discussions.
This work was supported by the U.S.\ Department of Energy, Office of Science, Basic Energy Sciences, Materials Sciences and Engineering Division. K.\ W. and R.\ W.\ acknowledge support from the Swiss National Science Foundation through the Postdoc Mobility program. Work at Brookhaven is supported by the Office of Basic Energy Sciences, U.S. Department of Energy under Contract No. DE-SC0012704.
\end{acknowledgments}

\end{document}


\title{Supplementary Material}

\date{\today}

\maketitle
\onecolumngrid

\section{Anisotropic $H_{c2}$ in a single-component superconductor}
For a single-component superconductor, an anisotropy of the upper critical field $H_{c2}$ near $T_{c}$ relates (i) to the shape of the electronic Fermi surface and (ii) to the superconducting gap function (see e.g. Ref. \cite{Kogan2012}) through

\begin{align}\label{eq:Gamma-swave}
\Gamma \equiv \frac{H_{c2}(0)}{H_{c2}(\pi/2)} = \bigg(\frac{\langle\phi^{2}(\vec{k})v_{\sss F}(0)^{2}\rangle}{\langle\phi^{2}(\vec{k})v_{\sss F}(\pi/2)^{2}\rangle} \bigg)^{1/2}.
\end{align}
Here, $\langle\boldsymbol{\cdot}\rangle$ denotes the average over the Fermi surface, $\phi(\vec{k})$ is the normalized gap function with $\langle\phi^2(\vec{k})\rangle=1$ and the Fermi momentum $\vec{k}$, and $v_{\sss F}(0)$, $v_{\sss F}(\pi/2)$ are the Fermi velocities along the maximal and minimal $H_{c2}$ directions (chosen along $\theta = 0$ and $\theta = \pi/2$ respectively). For
arbitrary magnetic field direction, the angular-dependent upper critical field near $T_{c}$ is
\begin{align}
H_{c2}(\theta)=\frac{H_{c2}(0)}{\sqrt{\cos^{2}\theta+\Gamma^{2}\sin^{2}\theta}}.\label{eq:HC2-one}
\end{align}

Given the absence of normal-state anisotropy in $\SBS$, it is rather unlikely that this scenario is at the origin of the anisotropy observed in $H_{c2}$, $\Gamma_{\mathrm{exp}} \approx 3.5$.

\section{Anisotropic $H_{c2}$ in a two-component superconductor}

In an earlier work, Venderbos \emph{et al.} \cite{Venderbos2016} have identified another mechanism leading to an anisotropy in $H_{c2}$. In fact, in two-componet nematic superconductors, the two order parameter components couple differently to a so-called pinning field. As a consequence a large anisotropy can arise even when normal-state properties appear isotropic. In order to illustrate this we follow an approach laid out by Venderbos and co-workers and describe a two-component order parameter $\vec{\eta} = (\eta_{1},\eta_{2})$. Treated within a Ginzburg-Landau (GL) approach, the free energy (near $T_c$) reads
\begin{align}\label{}
f_{\text{GL}}=
   (T &- T_{c0}) (|\eta_{1}|^{2} + |\eta_{2}|^{2}) - \delta T_{c0} (|\eta_{1}|^{2} - |\eta_{2}|^{2}) + J_{1}(D_{i}\eta_{a})^{*}D_{i}\eta_{a} +J_{2}\varepsilon_{ij}\varepsilon_{ab}(D_{i}\eta_{a})^{*}D_{j}\eta_{b} + J_{3}(D_{z}\eta_{a})^{*}D_{z}\eta_{a} \\\nonumber
   &+ J_{4}[|D_{x}\eta_{1}|^{2} + |D_{y}\eta_{2}|^{2} -|D_{x}\eta_{2}|^{2} -|D_{y}\eta_{1}|^{2} +(D_{x}\eta_{1})^{*}D_{y}\eta_{2} +(D_{y}\eta_{1})^{*}D_{x}\eta_{2} +(D_{x}\eta_{2})^{*}D_{y}\eta_{1} +(D_{y}\eta_{2})^{*}D_{x}\eta_{1}]
\end{align}
With $T$ the system's temperature, $T_{c0}$ the critical temperature (in absence of a symmetry-breaking field $\delta$), $J_{1}$, $J_{2}$, $J_{3}$, $J_{4}$ phenomenological parameters for the gradient terms, $D_{j} = -i\partial_{j} - 2e A_{j}$ the gauge-invariant derivatives with $j=x,y$, $\vec{A}$ the electromagnetic vector potential, $\varepsilon_{ij}$ the totally antisymmetric Levi-Civita tensor, and $\delta$ uniaxial distortion, favoring one order parameter over the other. The star denotes complex conjugation. Without limiting the foregoing, we assume $\delta > 0$ and $\mu > 0$ in the following. As will become apparent shortly, the ratio $\mu = J_{4}/J_{1}$ plays an important role in determining the anisotropy in $H_{c2}$.

In this form, a finite value of the order parameter $\eta_{1}$ appears at the (zero-field) critical temperature $T_{c} = T_{c0}(1+\delta)$. Dividing the above equation by $T_{c}$ reduces it to a dimensionless form,
\begin{align}\label{}
   (t &- 1)|\eta_{1}|^{2} + (t-1+2\Delta)|\eta_{2}|^{2} + J(D_{i}\eta_{a})^{*}D_{i}\eta_{a} + K (D_{z}\eta_{a})^{*}D_{z}\eta_{a} \\\nonumber
   &+ \mu J[|D_{x}\eta_{1}|^{2} + |D_{y}\eta_{2}|^{2} -|D_{x}\eta_{2}|^{2} -|D_{y}\eta_{1}|^{2} +(D_{x}\eta_{1})^{*}D_{y}\eta_{2} +(D_{y}\eta_{1})^{*}D_{x}\eta_{2} +(D_{x}\eta_{2})^{*}D_{y}\eta_{1} +(D_{y}\eta_{2})^{*}D_{x}\eta_{1}],
\end{align}
where we have (i) assumed that the field is applied within the $xy$ plane, (ii) neglected modulations along the field direction, (iii) introduced the dimensionless quantities $t = T/T_{c}$, $J = J_{1}/T_{c}$, $K = J_{3}/T_{c}$, and $\Delta = \delta/(1+\delta)$. Following the derivation in the Supplementary Materials of Ref.\ [\onlinecite{Venderbos2016}], the associated GL equations take the form
\begin{align}\label{eq:general}
   \frac{1}{K}\left(\begin{array}{c}
(1-t) \eta_{1}\\
(1-t - 2\Delta)\eta_{2}
\end{array}\right)
= \bigg[2e H \left((\hat{P}^{2}+\omega^{2}\hat{X}^{2})\vec{\tau}_{0}-\mu \omega^{2} \hat{X}^{2}(\vec{\tau}_{3}\cos2\theta+\vec{\tau}_{1}\sin2\theta)\right)\bigg]
\left(\begin{array}{c}
\eta_{1}\\
\eta_{2}
\end{array}\right)
\end{align}
with $\hat{X} = [\sin(\theta) D_{x} - \cos(\theta) D_{y}]/\sqrt{2eH}$, and $\hat{P} = D_{z}/\sqrt{2eH}$ denoting effective position and momentum operators (satisfying canonical commutation relations), $\omega^{2} = J/K$, and $\vec{\tau}_{i}$ Pauli matrices. Whereas the full angular dependence of $H_{c2}$ is obtained from a solution of the above equation (which is usually treated with perturbative methods), here we shall consider the special angles $\theta = 0$ and $\theta = \pi/2$. For the first case, $\theta = 0$ the above matrix equation reduces to the diagonal form
\begin{align}\label{}
   \frac{1}{K}\left(\begin{array}{c}
(1-t) \eta_{1}\\
(1-t - 2\Delta)\eta_{2}
\end{array}\right)
= 2eH\left(\begin{array}{cc}
\hat{P}^{2}+\omega_{-}^{2}\hat{X}^{2} & 0\\
0 & \hat{P}^{2}+\omega_{+}^{2}\hat{X}^{2}
\end{array}\right)
\left(\begin{array}{c}
\eta_{1}\\
\eta_{2}
\end{array}\right),
\end{align}
with $\omega_{\pm}^{2} = (J/K)(1\pm\mu)$. At $H_{c2}$, the system realizes the lowest Landau level (harmonic mode) in $\vec{\eta}$, that is either of the eigenstates $\vec{\eta} = (|\omega_{-}\rangle, 0)$ or $\vec{\eta} = (0, |\omega_{+}\rangle)$, where $(\hat{P}^{2}+\omega_{\pm}^{2}\hat{X}^{2})|\omega_{\pm}\rangle=\frac{1}{2}\omega_{\pm}|\omega_{\pm}\rangle$ represents the lowest harmonic mode with frequency $\omega_{\pm}$. As a consequence, $H_{c2}$ is determined by the appearance of the first instability, i.e.
\begin{align}\label{}
   H_{c2}(0) = \frac{1}{\sqrt{J K} e} \max \Big[\frac{1-t}{\sqrt{1-\mu}}, \frac{1-t - 2\Delta}{\sqrt{1+\mu}}\Big].
\end{align}
For the field applied along the orthogonal direction $\theta = \pi/2$, \eq \eqref{eq:general} also becomes diagonal
\begin{align}\label{}
   \frac{1}{K}\left(\begin{array}{c}
(1-t) \eta_{1}\\
(1-t - 2\Delta)\eta_{2}
\end{array}\right)
= 2eH\left(\begin{array}{cc}
\hat{P}^{2}+\omega_{+}^{2}\hat{X}^{2} & 0\\
0 & \hat{P}^{2}+\omega_{-}^{2}\hat{X}^{2}
\end{array}\right)
\left(\begin{array}{c}
\eta_{1}\\
\eta_{2}
\end{array}\right),
\end{align}
and we find the upper critical field from
\begin{align}\label{}
   H_{c2}(\pi/2) = \frac{1}{\sqrt{J K} e} \max \Big[\frac{1-t}{\sqrt{1+\mu}}, \frac{1-t - 2\Delta}{\sqrt{1-\mu}}\Big].
\end{align}
For the specific choice of signs ($\Delta > 0$, and $\mu > 0$) we then find after some calculus
\begin{align}\label{}
   H_{c2}(0) &= \frac{1}{\sqrt{JK}e}\frac{1-t}{\sqrt{1-\mu}} &
   H_{c2}(\pi/2) &= \frac{1}{\sqrt{JK}e} \left\{\begin{array}{c l}
\displaystyle \frac{1-t}{\sqrt{1+\mu}} & \text{for } \Delta > \Delta_{c}\\[1.5em]
\displaystyle \frac{1-t-2\Delta}{\sqrt{1-\mu}} & \text{for } \Delta < \Delta_{c}
\end{array}\right.
\end{align}
with $\Delta_{c}(\mu,t) = \{1-[(1-\mu)/(1+\mu)]^{1/2}\}(1-t)/2$. Independent of the precise angular dependence of $H_{c2}$, it follows from this analysis, that the anisotropy ratio $\Gamma$ takes the simple form
\begin{align}\label{}
   \Gamma_{>} &= \sqrt{\frac{1+\mu}{1-\mu}} \quad \text{for } \Delta > \Delta_{c}&
   \text{or} &&
   \Gamma_{<} &= 1 - \frac{2\Delta}{(1-t) - 2\Delta} \quad \text{for } \Delta < \Delta_{c}.
\end{align}
In the first case of a strong symmetry-breaking term, the anisotropy is independent of temperature and $\Delta$. In terms of the parameter, at $H_{c2}$ it assumes the form $\vec{\eta} = (\eta_{1},0)$ along both field directions. In the second case, of weak symmetry-breaking the anisotropy is independent of $\mu$. Here the order parameter takes the form $\vec{\eta} = (\eta_{1},0)$ when the field is along $\theta = 0$, and $\vec{\eta} = (0,\eta_{2})$ when the field is directed along $\theta = \pi/2$. In the analyses provided in Ref.\ \cite{Venderbos2016}---where the full angular dependence is considered in certain limiting cases---these anisotropy ratios are recovered.